# The NEFOCAST System for Detection and Estimation of Rainfall Fields by the Opportunistic Use of Broadcast Satellite Signals

Filippo Giannetti, Marco Moretti, Ruggero Reggiannini
*Dipartimento di Ingegneria dell'Informazione*
*University of Pisa*
Pisa, Italy
{filippo.giannetti, marco.moretti, ruggero.reggiannini}
@unipi.it

Attilio Vaccaro
*M.B.I. S.r.l.*
Pisa, Italy
avaccaro@mbigroup.it

*Abstract* — In this paper we present results from the NEFOCAST project, funded by the Tuscany Region, aiming at detecting and estimating rainfall fields from the opportunistic use of the rain-induced excess attenuation incurred in the downlink channel by a commercial DVB satellite signal. The attenuation is estimated by reverse-engineering the effects of the various propagation phenomena affecting the received signal, among which, in first place, the perturbations factors affecting geostationary orbits, such as the gravitational attraction from the moon and the sun and the inhomogeneity in Earth mass distribution and, secondly, the small-scale irregularities in the atmospheric refractive index, causing rapid fluctuations in signal amplitude. The latter impairments, in particular, even if periodically counteracted by correction maneuvers, may give rise to significant departures of the actual satellite position from the nominal orbit. A further problem to deal with is the daily and seasonal random fluctuation of the rain height and altitude/size of the associated melting layer. All of the above issues lead to non-negligible random deviations from the dry nominal downlink attenuation, that can be misinterpreted as rain events. In this paper we show how to counteract these issues by employing two differentially-configured Kalman filters designed to track slow and fast changes of the received signal-to-noise ratio, so that the rain events can be reliably detected and the relevant rainfall rate estimated.

*Keywords: Rain attenuation, Kalman filter, satellite communications, opportunistic rain rate estimation, nowcasting.*

I. INTRODUCTION

Monitoring of precipitations over a regional territory is an objective of major concern for public administrators in order to guarantee an adequate level of security for people living and operating on the territory [1], [2]. In particular, great importance is given to the availability of rain maps featuring *i*) good accuracy in the measurement of rainfall rate, *ii*) space-time completeness and continuity, and *iii*) negligible delays in the esti-mates of rain, so as to be able to predict from the above precipitation maps the arrival of severe meteorological events and adopt timely measures to prevent/reduce the related risk. The methods used to obtain a rain map are traditionally based, alternatively or jointly, on: a network of rain gauges [3], deployment of weather radars [4], satellite remote sensing in the optical or infrared band [5]. Each of the cited methods entails a more or less satisfactory degree of compliance with the requirements listed above. In [6] the advantages and limitations inherent in the use of these techniques are discussed. Independent of the approach taken, a terrestrial operational center is also needed for processing, control and alert functions.

More recently, in addition to the techniques above mentioned, it has been demonstrated [7]-[12] that rain fields can also be estimated by exploiting the presence of "signals of opportunity" generated by communication systems. In particular, for satellite systems, these signals are transmitted in the microwave region, usually in the Ku range or higher, and suffer from a *significant excess attenuation* in the presence of rain. In other terms, they exhibit a measurable reduction in the signal-to-noise ratio (SNR) with respect to dry operation, and this reduction is uniquely related to the intensity of rain. Moreover, they can be received by a potentially huge number of domestic devices, widely spread over the whole territory (at least within its anthropized part). This prospective "opportunistic" approach to the construction of rain maps has spurred the NEFOCAST project, funded by the Tuscany Region from 2016. A significant body of results on the project progress are already available in [6], [13]-[17]. The NEFOCAST project is based on the use of interactive domestic terminals capable of continuously measuring the received SNR and relaying the data to the operational center.

The problem of estimating the rainfall rate from a measure of excess signal attenuation or SNR reduction on a satellite link, has already received considerable attention in the literature (for a representative selection of references see [9]-[12]), and its points of force and weakness have been investigated. One of the most critical problems, at the moment not yet satisfactorily resolved, is how to choose the SNR reference level corresponding to the absence of rain (a condition referred to as *dry* in the sequel), against which the rain-induced SNR variations should be measured. As a matter of fact, the downlink dry signal level is far from being exactly constant, due to various phenomena, notably the apparent movement of the satellite under the attraction of the sun and the moon.

In spite of the above drawbacks, the method has however some undoubted advantages compared to conventional techniques, notably, it is not subject to significant measurement delays, unlike tipping bucket rain gauges, affected by delays of several minutes, especially in response to small rainfall rates. Furthermore, the geographic capillarity achievable by this technique seems prospectively insuperable, if only each of the domestic receivers in use by the population could be provided with the measurement and communication capability of NEFOCAST terminals.

In this paper we present the approach pursued in the NEFOCAST project, showing that it is capable to provide a satisfactory solution to most of the issues outlined above. Specifically, we propose a novel processing architecture featuring two Kalman Filters (KFs) which allow an accurate tracking of SNR fluctuations in both dry and wet conditions, and also a reliable detection of the rain start/stop epochs. One of the two KF, called *slow tracker* (ST), has the function of following the slow variations of SNR, with 24-h period, due to station-keeping control manoeuvers, while eliminating all faster components (scintillation, rain) in the received SNR. In other terms, its task is to provide a clean time-varying dry reference SNR against which to measure the SNR changes induced by rain. The other KF, termed *fast* tracker (FT), has the task of smoothing out the very fast SNR fluctuations due to ionospheric scintillation, but must be able to follow, with negligible errors and delay, the variations of SNR due to rain, whose typical persistence time (order of several minutes or more) is much longer than the fluctuations of scintillation noise, but much shorter than the period of station-keeping effects.

Taking the difference between the outputs of the slow and fast trackers permits to achieve a twofold objective: first, we can detect the presence of rain, whenever the above difference exceeds a certain threshold; second, the difference can be mapped uniquely onto an estimate of rainfall rate. This approach seems to be advantageous compared to the one using a single sliding-window filter to provide the dry reference SNR, as suggested in [11], since the KF, thanks to its intrinsic iterative prediction-correction mode of operation, is subject to a tracking delay smaller than that affecting the sliding-window filter (this being exactly equal to half the window length). Moreover, the NEFOCAST rain retrieval algorithm features an additional advantage compared to previous published procedures: in NEFOCAST, calculation of the specific rainfall attenuation is carried out using measurements (or very-short-term forecasts) of the 0°C isotherm height, a parameter supplied by the CNR-IBIMET/LAMMA weather forecast service with a 6-hour rate for the whole Tuscany region. This allows to achieve a greater precision compared to other approaches proposed in the literature. For instance, in [11] the 0°C isotherm height is identified by making an *a priori* guess on the basis of the available statistics for the time (month, day, hour) and place where the rain estimate is made. To further improve rain estimation accuracy in NEFOCAST, a method is also implemented (and illustrated in Sect. IV) to model the different absorbing properties of the melting layer with respect to the full liquid rain volume.

The paper is organized as follows: in Sect. II the structure of the NEFOCAST sensor network is outlined. Sect. III illustrates the proposed procedure for estimating the rain rate in ideal conditions. In Sect. IV we compute the SNR for a satellite link and, in the case of a rain event, outline the SNR dependence on rain attenuation. Sect. V describes the satellite link geometry and presents a model for relating the rain attenuation to the rain rate. In Sect. VI we briefly review the impairments due to propagation anomalies and antenna mispointing affecting the satellite link. In Sect. VII and Sec. VIII we illustrate the scheme for detection and estimation of rain-induced SNR variations in the presence of the impairments of a real system, based on the double slow-fast KF. Numerical results are presented in Sect. IX. Finally, conclusions follow in Sect. X.

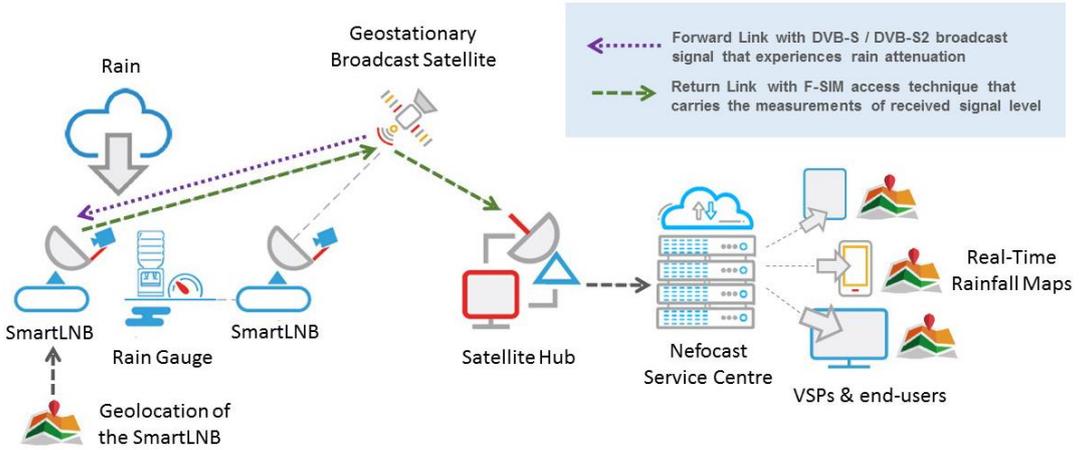

Fig. 1. NEFOCAST experimental network for real-time wide-area high-spatial resolution rain-rate measurement.

## II. THE NEFOCAST PROJECT

NEFOCAST is a research project, funded by the government of the Tuscany Region (Italy), aimed at assessing the feasibility of a satellite-based system for real-time construction of rain maps over the regional territory [13]-[15]. The main idea is to make opportunistic use of broadcast satellite signals related to domestic services such as DVB-S/S2 to estimate the intensity of rain events by measuring the satellite signal attenuation at the receiver. The large number of terminals and the possibility of having real time measurements are key elements for the implementation of an efficient nowcasting platform.

A sketch of the concept of the NEFOCAST system is illustrated in Fig. 1: a large number of geolocated satellite receivers (interactive satellite terminals, ISTs) act as rains sensors and send their measurements to a satellite hub on the return link. The IST employed in the NEFOCAST project, which serves both as weather sensor and modem, is an innovative two-way (i.e., transmit/receive), low-cost, small-size, low-power device named SmartLNB (Smart low noise block) [16]. Figure 2 shows the parabolic antenna installed with its own SmartLNB at the site of the University of Pisa. The observations of the ISTs are integrated with the measurements from a certain number of tipping bucket rain gauge (TBRG) sensors, used as reference benchmarks for validation. The satellite hub redirects the data to the NEFOCAST service center (NSC), which is able to collect spatially accurate information from a wide network of terminals and TBRG sensors to generate and validate rain field maps that can be shared with a number of value-added service providers (VSPs).

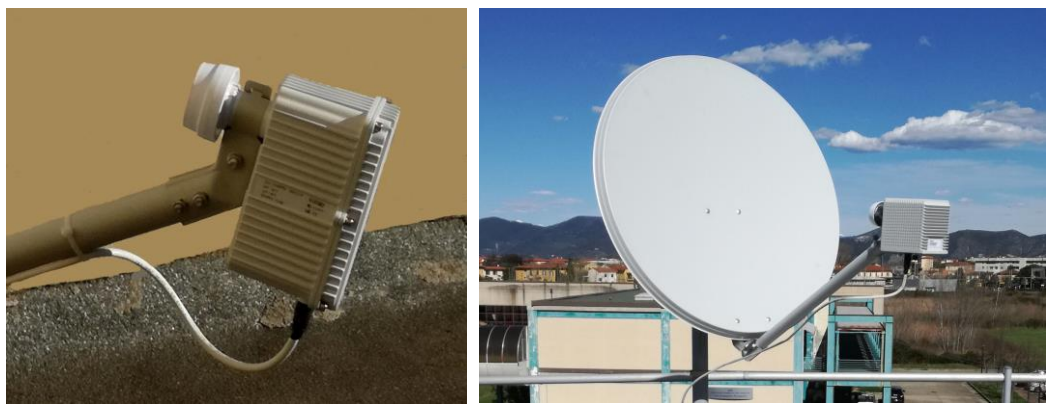

Fig. 2. *Left*: SmartLNB device. *Right*: 75 cm-parabolic antenna of the NEFOCAST station at the University of Pisa (NEFOCAST-ITA-PI-003X), with a SmartLNB.

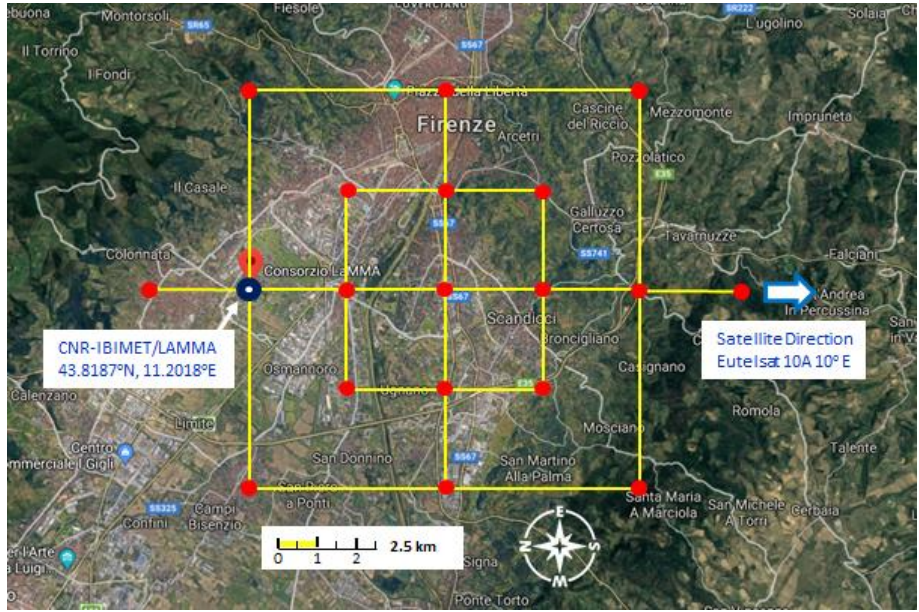

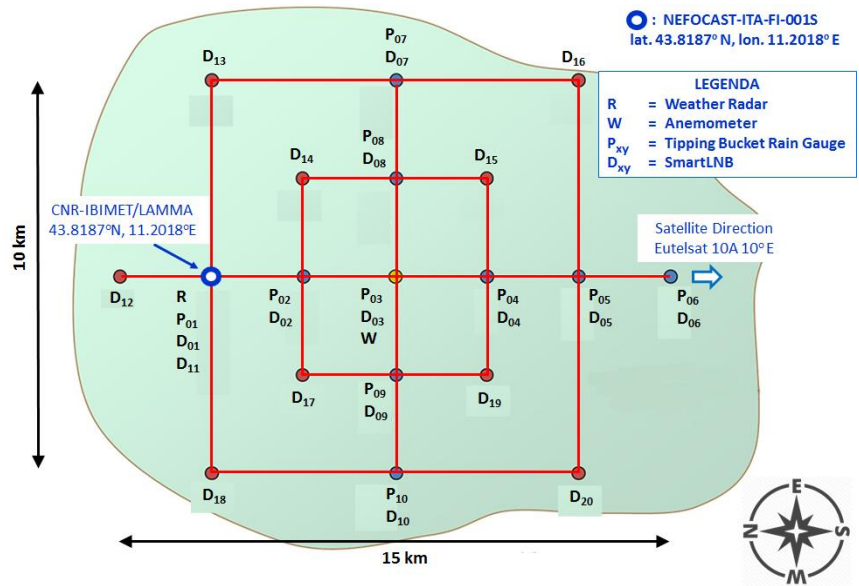

Fig. 3.   *Top:* satellite map of the metropolitan area of Florence around CNR-IBIMET/LAMMA (ground station NEFOCAST-ITA-FI-001S, 43.8187°N, 11.2018°E) portraying the locations on the 15 km × 10 km grid where the various sensors have been deployed (circular marks). *Bottom*: positions on the grid of the various type of sensors.

The NEFOCAST project has been organized in two different phases: 1) a *research phase*, where SmartLNBs measurements have been compared with the readings of TBRG sensors to develop an efficient algorithm for estimating rainfall rate from measurement of channel attenuation (this activity is now completed and its results are detailed in the next sections); 2) a *validation phase*, where the algorithm is tested over the metropolitan area of the city of Florence and the rain maps obtained by a network of several SmartLNBs are compared with the measurements of a set of TBRG sensors, anemometers and a weather radar. Figure 3 shows the map of the deployment of the SmartLNBs and TBRG sensors over the metropolitan area of Florence.

The main parameters of the NEFOCAST bidirectional network are summarized in Table I. Specifically, the forward-link (FL), i.e., satellite-to-ground, signal received by the SmartLNBs employs standard DVB-S2 modulation. The return-link (RL), i.e., ground-to-satellite, signal generated by the SmartLNBs uses Spread Spectrum Aloha (SSA) as modulation/medium access technique, implemented by the F-SIM (Fixed Interactive Multi-media Services) protocol [17].

## III. PROCEDURE FOR RAIN RATE RETRIEVAL

A key parameter for estimating the rain rate is the ratio $E_s/N_0$ (signal-to-noise ratio, or SNR, for short) measured at the ground terminal, wherein the numerator $E_s$ represents the average radiofrequency received energy (in J) within the time interval of one information-bearing symbol, and the denominator $N_0$ is the one-sided power spectral density of the additive white Gaussian noise (in W/Hz) affecting the received signal. In the presence of a rain event, the procedure for the estimation of the rainfall rate based upon the measured SNR can be summarized in four steps:

1. Obtain the reference value of the SNR in dry conditions, $[E_s/N_0(t)]^{(dry)}$;
2. Measure the actual SNR, $[E_s/N_0(t)]^{(wet)}$, whose value depends on the intensity of rain;
3. Extrapolate the attenuation of the SNR due to the rain, $L_{\text{rain}}(t)$;
4. Derive an estimate of the rain rate $\hat{R}(t)$ from $L_{\text{rain}}(t)$.

TABLE I. SUMMARY OF NEFOCAST SYSTEM MAIN PARAMETERS

| Feature / Item | Name / Value |
|---|---|
| Satellite name, orbital slot | Eutelsat 10A, 10° East |
| FL satellite EIRP | 48 dBW |
| FL frequency, polarization | 11.345 GHz, LVP |
| FL protocol, modulation, FEC rate | DVB-S2, QPSK, 4/5 |
| FL noise figure of SmartLNB | 0.2 dB @ 290 K |
| RL SmartLNB EIRP | From 15 dBW to 35 dBW |
| RL frequency, polarization | 14.216 GHz, LHP |
| RL protocol | SSA F-SIM |
| RL figure of merit G/T | 4 dB/K |

Unfortunately, as we will see in Sect. VI, in a *real* system there are several impairments that affect this procedure, which, if not taken care of, compromise the accuracy of the resulting estimate. In the following sections we first address in detail all the aspects related to ideal rain rate retrieval, then we analyze the impairments and finally propose a solution based on a double-KF architecture.

IV. SNR FOR A SATELLITE LINK AND ITS DEPENDENCE ON RAIN ATTENUATION

*SNR in Dry Conditions.* In case of *dry* conditions, i.e., in the absence of rain, the (dimensionless) SNR can be expressed as (notice that the SNR is actually a time-varying link quality metric due to the many impairments outlined in Sect. VI) [18]

$$\left[\frac{E_s}{N_0}(t)\right]^{(\text{dry})} \triangleq \frac{\frac{\Phi G_R \lambda^2}{4\pi R_s \text{k}}}{L_{\text{atm}} L_{\text{cloud}} \left[\frac{T_c}{L_{\text{atm}} L_{\text{cloud}}} + T_m \left(1 - \frac{1}{L_{\text{atm}} L_{\text{cloud}}}\right) + T_g + T_{\text{rx}}\right]}, \quad (1)$$

where $\Phi$ is the signal power flux density (in W/m$^2$) at the receiving antenna input, $G_R$ is the receiving antenna gain (dimensionless), $\lambda$ is the carrier wavelength (in m), $R_s$ is the symbol rate (in s$^{-1}$), $L_{\text{atm}}$ is the atmospheric attenuation due to water vapor absorption and other gaseous effects (dimensionless), $L_{\text{cloud}}$ is the attenuation due to clouds (dimensionless), k is the Boltzmann constant (in J/Hz) and $T_c$, $T_m$, $T_g$, $T_{rx}$ are the noise temperatures of cosmos, meteorological formations (clouds, etc.), environment surrounding the antenna and receiver hardware, respectively (all in K).

*SNR in Wet Conditions.* In case of *wet* conditions, i.e., in the presence of rain, the signal experiences an *additional* fast time-varying attenuation $L_{rain}(t)$ (dimensionless) and accordingly the SNR in (1) becomes (due to its role in the evaluation of the rain rate, the time-dependence of the rain attenuation is here highlighted)

$$\left[\frac{E_s}{N_0}(t)\right]^{(wet)} = \frac{\frac{\Phi G_R \lambda^2}{4\pi R_s k}}{L_{atm} L_{cloud} L_{rain}(t) \left[\frac{T_c}{L_{atm} L_{cloud} L_{rain}(t)} + T_m \left(1 - \frac{1}{L_{atm} L_{cloud} L_{rain}(t)}\right) + T_g + T_{rx}\right]} \quad (2)$$

The algorithm for the estimation of the rain rate requires knowledge of the rain attenuation $L_{rain}(t)$ on the satellite downlink, which from manipulation of (1) and (2) can be formally expressed as follows:

$$L_{rain}(t) = \frac{\left[(E_s/N_0)(t)\right]^{(dry)}}{\left[(E_s/N_0)(t)\right]^{(wet)}} (1-\xi) + \xi, \quad (3)$$

where

$$\xi \triangleq \frac{T_m - T_c}{L_{atm}(T_m + T_g + T_{rx})}. \quad (4)$$

Equation (3) represents the basic tool we use for the evaluation of $L_{rain}(t)$, and it requires in principle the availability of both the dry and wet SNRs at any instant *t* during a rain event. In practice, however, only the wet SNR is available (measurable) during a precipitation, and therefore only a presumed, fictitious value for the dry baseline SNR can in fact be used in (3). Estimation of the dry baseline SNR in rainy conditions is an open research issue and is currently being investigated by the authors. The results presented in this paper are obtained by replacing the dry SNR in (3), during rainy periods, with the last value of dry SNR measured before the rain onset, and keeping this value constant for the entire duration of the rainy event. Further details of this procedure are given in Sect. VII. The numerical values used to compute (4) are recapped in Table 2. The values are either taken from the literature [18] or directly measured at the receive station ID NEFOCAST-ITA-PI-003X (Pisa, 43.7203° N, 10.3836° E, elevation angle $\theta_e \cong 40°$) and lead to $\xi = 0.799$.

TABLE II. SUMMARY OF THE MAIN PARAMETERS OF THE SATELLITE LINK

| Feature / Item | Name / Value |
|---|---|
| $L_{atm}$ | 0.09 dB |
| $T_m$ | 275 K |
| $T_c$ | 2.78 K |
| $T_g$ | 45 K |
| $T_{rx}$ | 13.67 K |

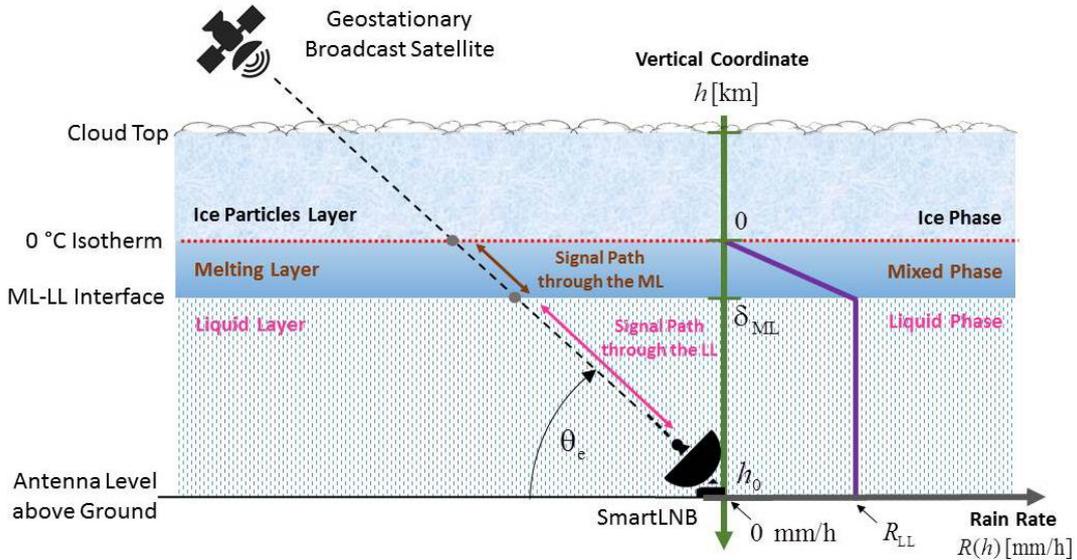

Fig. 4. *Left*: downlink geometry with stratiform rain. *Right*: rain rate (mm/h) vs. vertical coordinate (km).

V. LINK GEOMETRY AND MODEL FOR RAIN ATTENUATION

To evaluate the rain attenuation in a satellite link we make the simplified assumption of *stratiform* precipitation, characterized by a clear separation, at the height $h_0$ of the 0 °C isotherm, between the upper layer, named the "ice particles layer" (IPL), made of frozen dry particles, and the lower layers in which the ice particles melt into raindrops. Significant excess attenuation of the signal occurs only in the lower layers due to the presence of liquid-phase precipitation. These layers are structured as follows: the superior one is named "melting layer" (ML) and contains a combination of ice and rain, while the inferior is named "liquid layer" (LL) and contains only rain.

Figure 4 illustrates the geometry of the radio link and represents the various precipitation layers. To simplify the notation in the upcoming equations, the reference system has been chosen to set the zero quota at the level of 0 °C isotherm and the vertical coordinate is oriented to increase in the direction of the ground, so that at the ground antenna level the vertical coordinate is $h = h_0$. In the remainder of the paper the vertical coordinate is expressed in km and all the variables expressing a value of rain rate are in mm/h. In the LL, the time-varying rainfall rate $R_{LL}(t)$ is assumed independent of the vertical coordinate, while in the ML we assume that the liquid rain fraction varies linearly with height [19], passing from 0 at the 0 °C isotherm level, to the full liquid rate $R_{LL}(t)$ at the lower edge of the ML. Therefore, denoting with $R(h;t)$ the (liquid) rain rate as a function of both the time and the vertical coordinate $h$, we have

$$R(h;t) \triangleq \begin{cases} \dfrac{R_{LL}(t)}{\delta_{ML}} h, & 0 \leq h \leq \delta_{ML} \\ R_{LL}(t), & \delta_{ML} < h \leq h_0 \end{cases}, \qquad (5)$$

where $\delta_{ML}$ (km) is the vertical extension of the melting layer, given by the difference between the 0 °C isotherm level and the full liquid rain height [20], and we assumed $\delta_{ML} < h_0$. We also assume that the specific rain attenuation $k(h;t)$ in dB/km is a time-varying function of the vertical coordinate $h$, related to the rainfall rate $R(h;t)$ according to the customary $k = \alpha R^b$ power law. By adapting the power law to our 2-layer (ML-LL) model, one gets

$$k(h;t) \triangleq \begin{cases} k_{ML}(h;t) \triangleq \alpha_{ML}\left[R(h;t)\right]^{\beta_{ML}} = \alpha_{ML}\left[\dfrac{R_{LL}(t)}{\delta_{ML}} h\right]^{\beta_{ML}}, & 0 \leq h \leq \delta_{ML} \\ k_{LL}(h;t) \triangleq \alpha_{LL}\left[R(h;t)\right]^{\beta_{LL}} = \alpha_{LL}\left[R_{LL}(t)\right]^{\beta_{LL}}, & \delta_{ML} < h \leq h_0 \end{cases}. \quad (6)$$

The coefficients in the LL are supposed to be independent of $h$ and, for the NEFOCAST FL frequency (11.345 GHz), take on the values $\alpha_{LL} = 0.0153$ and $\beta_{LL} = 1.2531$, according to the results mentioned in [15], following a massive experimental campaign in central Italy. As for the ML, we assume a different set of coefficients, namely $\alpha_{ML} = 0.0914$ and $\beta_{ML} = 1.1068$ as in [21]. In the literature [19] there are also slightly different values proposed for the coefficients, but their choice would not impact significantly the results presented in the sequel. In all cases, $\alpha_{ML}$ and $\beta_{ML}$ are considerably larger than the coefficients relative to the LL, implying that the specific attenuation of the ML largely exceeds that of the LL. We now proceed calculating the overall rain attenuation at ground level.

*Rain Attenuation in the ML.* The elementary contribution to rain attenuation (in dB) introduced by an elementary layer of the ML at the vertical coordinate $h$, with thickness $dh$, on a propagation path slanted by an angle $\theta_e$ (i.e., the antenna elevation angle at the ground terminal) having length $d\ell \triangleq dh/\sin\theta_e$ is

$$dL_{rain}(h;t)\big|_{dB} \triangleq k_{ML}(h;t)\,d\ell = \frac{\alpha_{ML}\left[\dfrac{R_{LL}(t)}{\delta_{ML}}h\right]^{\beta_{ML}}}{\sin\theta_e}\,dh \text{ (dB)}, \quad 0 \leq h \leq \delta_{ML}. \qquad (7)$$

The total rain attenuation (in dB) introduced by the ML is thus given by the integration of (7) on the whole ML thickness

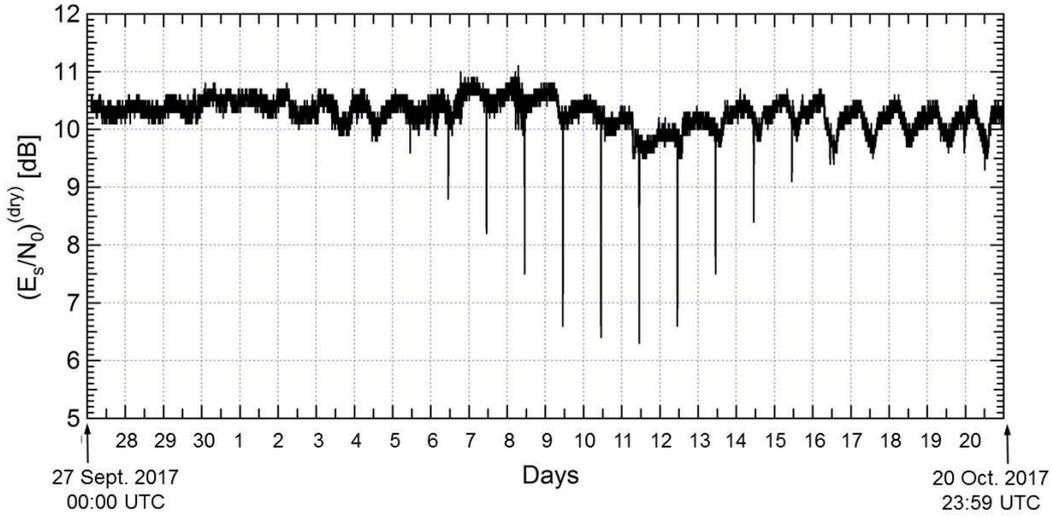

Fig. 5. Record of received dry $[(E_s/N_0)(t)]^{(dry)}$ (dB) measured by the SmartLNB of station NEFOCAST-ITA-PI-003X (Pisa, 43.7203° N, 10.3836° E), showing 24 h-periodic fluctuations due to GEO satellite orbit perturbations and a series of deep notches from 5 to 16 October 2017, caused by sun transit behind the satellite (Eutelsat 10A at 10° E) around local noon (about 11:00 UTC).

$$L_{ML}(t)\big|_{dB} \triangleq \int_0^{\delta_{ML}} \frac{\alpha_{ML}\left[\dfrac{R_{LL}(t)}{\delta_{ML}}h\right]^{\beta_{ML}}}{\sin\theta_e}\,dh = \frac{\alpha_{ML}\left[R_{LL}(t)\right]^{\beta_{ML}}\delta_{ML}}{\sin\theta_e(\beta_{ML}+1)} \text{ (dB)}. \qquad (8)$$

From (8) it is seen that the entire ML is equivalent, as far as the attenuation is concerned, to a LL characterized by the rainfall rate $R_{LL}(t)$, with coefficients $\alpha_{ML}$, $\beta_{ML}$ and an equivalent vertical extension $\delta_{eq} \triangleq \delta_{ML}/(\beta_{ML}+1)$. For instance, letting $\delta_{ML} = 0.5$ km, the equivalent thickness turns lout $\delta_{eq} \cong 0.24$ km, i.e., less than half the real value. On the other hand, if the same coefficients $\alpha_{LL}$ and $\beta_{LL}$ of the LL could be applicable to the whole ML as well, to achieve the same attenuation the thickness of the ML should be scaled accordingly by $\alpha_{ML} / \alpha_{LL} \cdot (\beta_{LL} + 1) / (\beta_{ML} + 1) \cdot R_{LL}^{\beta_{ML} - \beta_{LL}} \cong 6.39 \cdot R_{LL}^{-0.1463}$. Letting $\delta_{ML} = 0.5$ km and $R_{LL} = 10$ mm/h, this would lead to an equivalent thickness as large as $\delta_{eq} \cong 2.3$ km, i.e., almost five times the real value.

*Rain Attenuation in the LL.* In the LL (i.e., under the ML) the rainfall rate remains constant w.r.t. the vertical coordinate $h$ and equal to $R_{LL}(t)$. Therefore, the rain attenuation (in dB) introduced by an elementary layer of the LL at vertical coordinate $h$ and having thickness $dh$ on a propagation path slanted by an angle $\theta_e$ and having length $d\ell \triangleq dh / \sin\theta_e$ is now

$$dL_{rain}(h;t)\big|_{dB} \triangleq k_{LL}(h;t)\, d\ell = \frac{\alpha_{LL}\left[R_{LL}(t)\right]^{\beta_{LL}}}{\sin\theta_e} dh \text{ (dB)},\ \delta_{ML} < h \leq h_0. \qquad (9)$$

The total rain attenuation (in dB) introduced by the LL is thus given by the integration of (9) on the whole LL thickness to yield

$$L_{LL}(t)\big|_{dB} \triangleq \int_{\delta_{ML}}^{h_0} \frac{\alpha_{LL}\left[R_{LL}(t)\right]^{\beta_{LL}}}{\sin\theta_e} dh = \frac{\alpha_{LL}\left[R_{LL}(t)\right]^{\beta_{LL}}}{\sin\theta_e}(h_0 - \delta_{ML}) \text{ (dB)} \qquad (10)$$

*Total Attenuation due to rain.* During a precipitation event the radio signal has to propagate through the IPL, the ML and the LL, and each layer is amenable to give its contribution to the total excess attenuation. However, since the contribution of the IPL is typically negligible, i.e., $L_{IPL}\big|_{dB} \simeq 0$ dB, the total excess attenuation due to a precipitation event reduces to

$$L_{rain}(t)\big|_{dB} \triangleq L_{ML}(t)\big|_{dB} + L_{LL}(t)\big|_{dB} \text{ (dB)}. \tag{11}$$

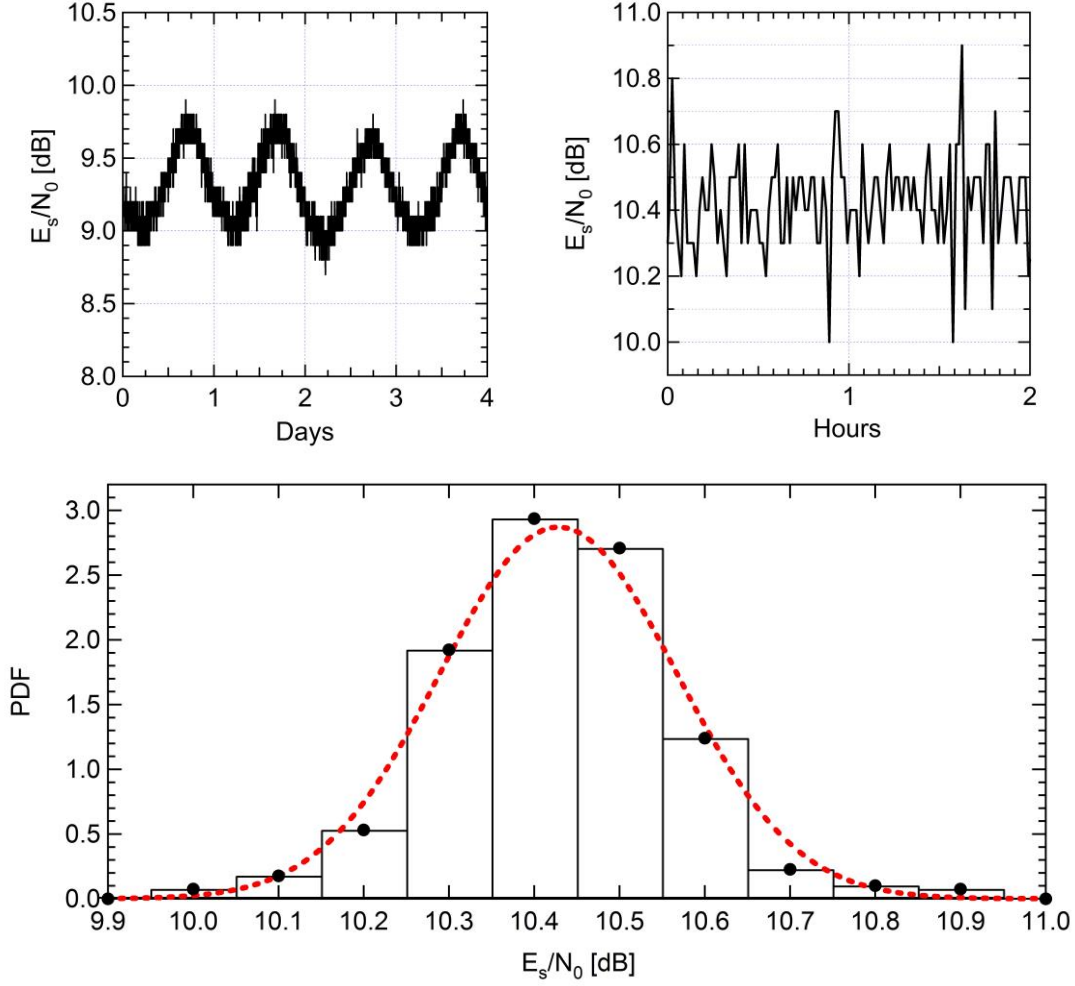

Fig. 6. Record of received dry SNR (dB) measured by the SmartLNB of station NEFOCAST-ITA-PI-003X (Pisa, 43.7203° N, 10.3836° E) with 1 min sampling interval and 0.1 dB amplitude resolution. *Top left*: large-scale (4 days) measurement showing the 24 h-periodic fluctuations due to GEO satellite orbit perturbations. *Top right*: small scale (2 hours) measurement showing the fast fluctuations due to tropospheric scintillation. *Bottom*: histogram of the probability density function of the scintillation noise and the relevant Gaussian fitting (dashed line); mean 10.428 dB, standard deviation 0.139 dB.

Recalling (10) and (12), one finally obtains the overall rain attenuation for the typical wet conditions as a function of the rainfall rate at ground $R_{LL}(t)$ and the other parameters, as follows

$$L_{rain}(t)\big|_{dB} \triangleq \overbrace{\alpha_{ML}\left[R_{LL}(t)\right]^{\beta_{ML}} \left[\frac{\delta_{ML}}{(\beta_{ML}+1)\sin\theta_e}\right]}^{\text{Melting layer (ML)}} + \overbrace{\alpha_{LL}\left[R_{LL}(t)\right]^{\beta_{LL}} \left(\frac{h_0 - \delta_{ML}}{\sin\theta_e}\right)}^{\text{Liquid layer (LL)}} \text{ (dB)}. \tag{12}$$

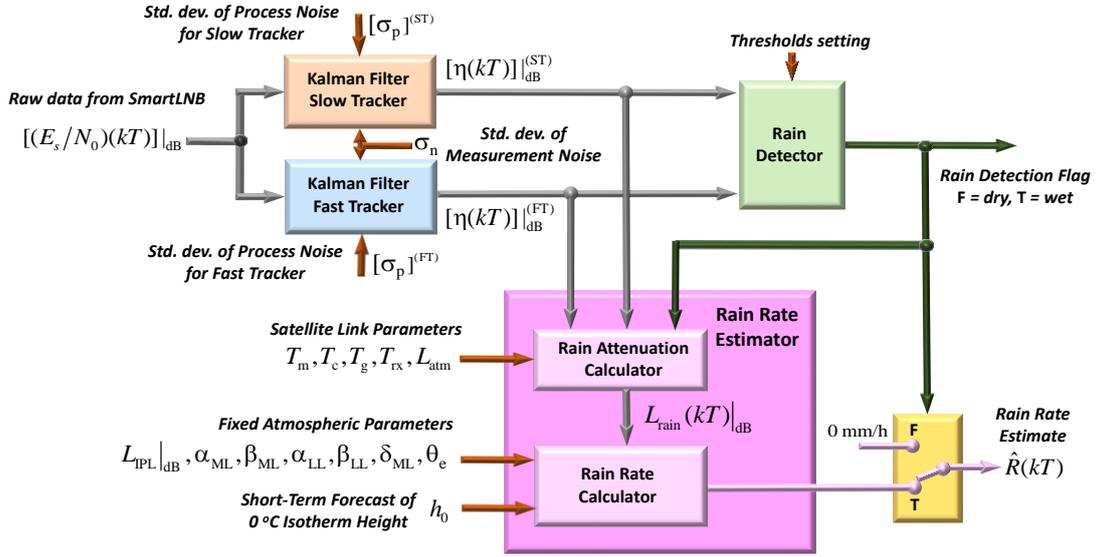

Fig. 7.  Block diagram of the NEFOCAST rain rate estimation algorithm based on a double Kalman filter.

Given a measured rain attenuation $L_{\text{rain}}(t)|_{\text{dB}}$, inversion of (12) yields an estimate of $R_{\text{LL}}(t)$. However, due to the transcendental nature of (12), this inversion calls for an iterative procedure or a tabulation.

Summarizing the above procedure, at any sampling instant $t_k = kT$, from the current measurement of the received SNR produced by the SmartLNB, the algorithm calculates the corresponding rain attenuation $L_{\text{rain}}(t_k)|_{\text{dB}}$ experienced by the radio signal and finally from (12) provides the estimate $\hat{R}(t_k)$ of the rain rate. In a practical implementation of the algorithm, the value of $h_0$ involved in (12) is obtained from models, such as for instance the WRF (Weather Research and Forecasting model). Furthermore, from analysis of (12) it is seen that rain rate estimates are quite sensitive to errors in the knowledge of $h_0$, and therefore the use of good-quality forecasts of $h_0$ is recommended. Instead, requirements about $\delta_{\text{ML}}$ are more relaxed. Actually, a specific investigation (not presented here due to lack of space) on the impact of a mis-match in the knowledge of $\delta_{ML}$ on rainfall estimation accuracy, showed that using in (12) values belonging to the interval $\delta_{ML} \approx 300 \div 800$ m (a typical range for temperate climates [20], [22]-[24]) yields rainfall rate estimates lying within ±8% with respect to the estimate obtained with fixed "average" value $\delta_{ML} = 500$ m.

VI. IMPAIRMENTS AFFECTING THE LEVEL OF A SIGNAL RECEIVED FROM A GEO SATELLITE

Even in dry conditions, the signal received from a satellite may be affected by many impairments [25] that cause amplitude fluctuations. An example is shown in Fig. 5, which plots the value of $[(E_s/N_0)(t)]^{(dry)}$ in the period 5-16 October 2017 for a specific SmartLNB. The presence of these impairments greatly affects the rain retrieval algorithm since the accuracy of the attenuation estimate in (3) depends on the fact that $[(E_s/N_0)(t)]^{(wet)}$ should differ from $[(E_s/N_0)(t)]^{(dry)}$ only because of the rain. A critical part of the NEFOCAST project has been identifying the main factors that perturb the measure in (3) so that specific countermeasures could be taken. These are the main impairments:

- *Scintillation fading.* Scintillation fading [26] denotes rapid fluctuations in signal amplitude caused by small-scale irregularities in the tropospheric refractive index. This effect is significant for frequencies above 10 GHz and grows with frequency. In Ku band, fluctuations are within ±0.5 dB and the period of scintillation fades varies from 1 to 10 s. Accordingly, the spectral width of the fluctuations in the Ku band is about 0.1 Hz [26]. These fluctuations are thus much faster than the rain events and even faster if compared to the long-term effects mentioned here below, and as such they can be smoothed out by a KF (Sect. VII).

- *Orbit perturbations.* In practice, a satellite is subject to many sources of perturbations that make it impossible to maintain its orbit perfectly stable. One of the main orbit perturbations is related to the gravitational effects of the moon and the sun that cause a progression of the orbit inclination [27]. These perturbations are periodically counteracted by means of orbit correction manoeuvers. The residual orbit inclination causes an apparent daily movement of the satellite in elevation and longitude, as seen from the ground station, along an 8-shaped path, with a 24 hours period [18]. The value of the receiving antenna gain towards the satellite is thus continuously changing and causes the daily nearly-periodic fluctuations visible in Fig. 5. These long-term signal fluctuations can be effectively tracked using a sufficiently slow KF (Sect. VII).

- *Other sources of long-term signal fluctuations.* Further slow variations of the received signal amplitude are caused by the drift of the longitudinal satellite orbital position or by beam bending caused by large-scale changes in the medium refractive index, due to the atmosphere temperature and humidity variations [28]. However, both of them have slow dynamics and can be tracked by the same KF (Sect. VII).

- *Sun transit.* Around the equinoxes, the Earth receiving stations pointing at a geostationary satellite are occasionally "blinded" by the sun's apparent passage behind the satellite. This phenomenon, which is referred to as "sun transit", lasts a few minutes daily, over a period of a few days and twice a year. During the transit, the sun's noise heavily interferes with the satellite FL signal and this leads to an increase of the antenna noise temperature, which causes severe SNR deterioration. However, the date, the time and the duration of any sun transit can be accurately predicted, and therefore such an issue can be ef-fectively dealt with with a proper management of the SmartLNB measurements. For instance, the last measured value of SNR prior to sun transit could be kept fixed for the whole duration of the fade event.

- *Changes of the transponder gain setting*. Occasionally, the satellite operator changes the transponder gain setting due to customer requests or other operational needs. If the change consists in a power reduction, the algorithm used to process the real-time data provided by the SmartLNBs will misinterpret it as the onset of a precipitative event. In this case, however, the signal fade simultaneously affects all sensors, and therefore it can be easily recognized and counteracted by a global management of the data at the processing center.

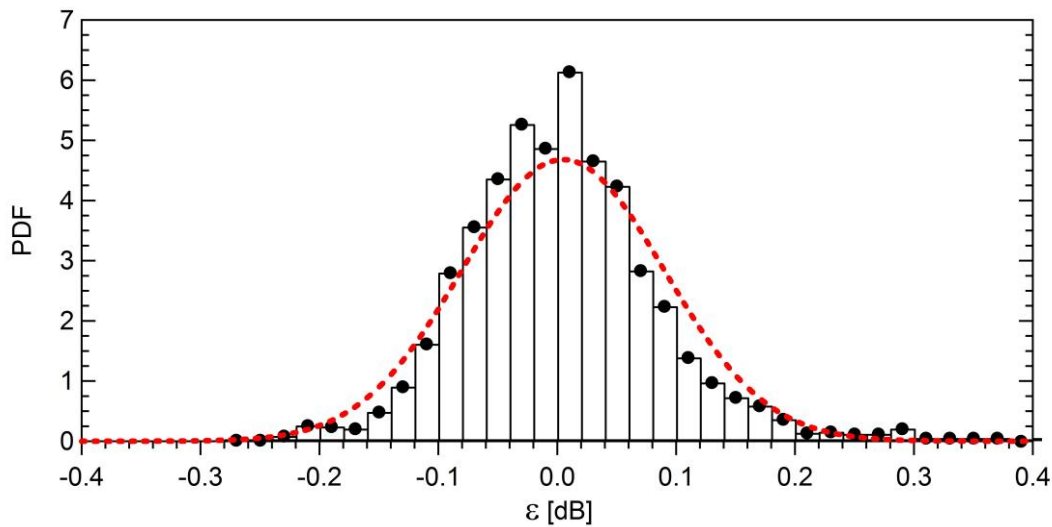

Fig. 8.  Histogram of the probability density function of the difference between ST and FT outputs (in dB) and the relevant Gaussian fitting (dashed line); mean 0.0049 dB, standard deviation 0.0852 dB.

The effects of some of these impairments are illustrated in Fig. 6, which shows the effect of scintillation noise and satellite orbit perturbation. The histogram of the probability density function of the scintillation noise in dry periods shows a good fitting with a Gaussian distribution. It is important that all parameters in the NEFOCAST algorithm are appropriately geared so as to minimize the probability that the above impairments are misinterpreted as the occurrence of a rain event.

## VII. Slow-Fast Differential Kalman Tracker

Moreover, in the design of the algorithm for retrieving the rain rate from the measured SNR there are two other important issues, which need dedicated signal processing:

1) In (3) the evaluation of $L_{\text{rain}}(t)$ requires the knowledge of both $[(E_s/N_0)(t)]^{(\text{dry})}$ and $[(E_s/N_0)(t)]^{(\text{wet})}$. Unfortunately, during a precipitation event the former is not available, since the SmartLNB only provides measurements of $[(E_s/N_0)(t)]^{(\text{wet})}$. Accordingly, since the reference value $[(E_s/N_0)(t)]^{(\text{dry})}$ is a time-variant variable, the algorithm needs $[(E_s/N_0)(t)]^{*(\text{dry})}$, a fictitious *a priori* guess of the values of the SNR in dry conditions at the time *t*.

2) The algorithm needs a simple and reliable procedure, based only on the observation of the current values of SNR, for real-time detection of both the *beginning* and the *end* of a precipitative event. Again, to perform this task it is necessary to have an estimate of the "dry" sequence $[(E_s/N_0)(t)]^{(\text{dry})}$ even during a precipitative event, in order to use it as baseline against which to compare the current measurements of $[(E_s/N_0)(t)]^{(\text{wet})}$ provided by the SmartLNB. In fact, the onset of a precipitation can be detected when the latter sequence deviates significantly (downwards) from the former, while the end of the precipitation is revealed by the SmartLNB output converging back towards the "dry" values.

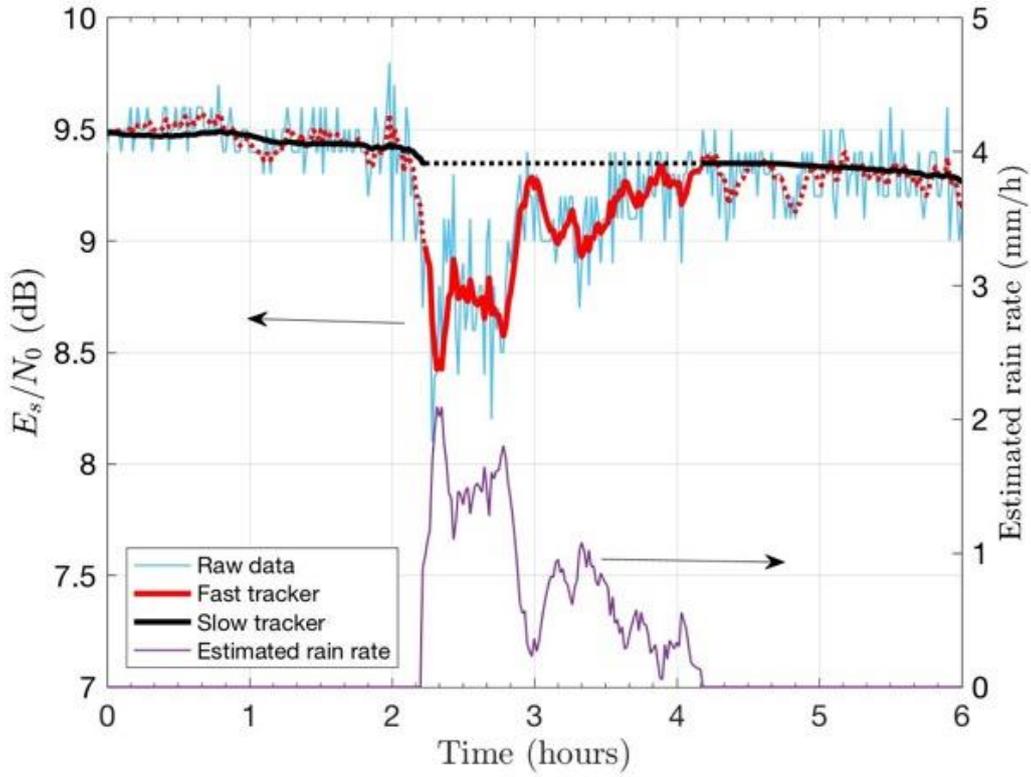

Fig. 9. *Top curves*: example showing the smoothing and tracking of the received $E_s/N_0$ samples (dB) using the NEFOCAST architecture with two Kalman filters in the presence of a rain event. *Bottom curve*: estimated rain rate (mm/h).

In order to solve the issues above, we note that SNR fluctuations during dry and wet conditions have a quite different behavior: specifically, during rain events, wet samples $[(E_s/N_0)(kT)]^{(wet)}$ have definitely faster dynamics (rate of variation) w.r.t. the dry ones. This suggests a method to distinguish between dry and wet conditions, consisting in the implementation of two estimators, having the structure of KFs and designed to process in parallel the SNR values measured by the SmartLNBs. The first estimator, hereinafter termed ST, is a simple two-state KF, tailored to track the very slow evolution of SNR in dry conditions (apparent satellite movements) and to be insensitive (or little sensitive) to scintillation noise and to the onset of rain events, whose time of persistence is typically in the order of several minutes or more. The second KF, described as FT, should be capable to accurately follow the faster-varying fluctuations of SNR induced by precipitations, removing only the scintillation noise.

The block diagram of the NEFOCAST architecture based on two KF is sketched in Fig. 7. The SmartLNB measures the received SNR and generates a stream of samples $[E_s/N_0(kT)]|_{dB}$ with fixed sampling period $T = 1 \min$. The two KFs, driven by the raw measurements of SNR, produce the time-discrete outputs $[\eta(kT)]|_{dB}^{(ST)}$, representing the estimate of $[(E_s/N_0)(kT)]|_{dB}^{(dry)}$ during dry periods, and $[\eta(kT)]|_{dB}^{(FT)}$, capable to track the fluctuations of $[(E_s/N_0)(kT)]^{(wet)}$ due to rain while smoothing out the scintillation noise.

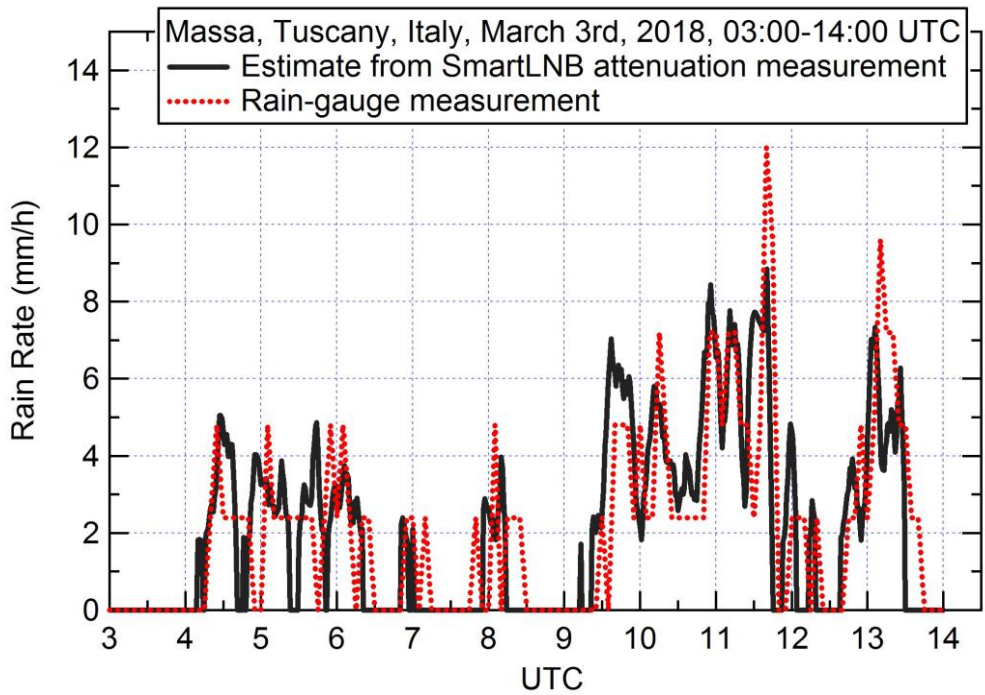

Fig. 10. Estimate of the rain rate obtained from SmartLNB measurements (solid line) made by station NEFOCAST-ITA-MS-001X (Massa, 44.0344º N, 10.1399º E), compared with conventional rain gauge measurements (dashed lines)

## VIII. RAIN DETECTION AND RAIN RATE ESTIMATION

In dry conditions the two KF trackers provide approximately the same filtered output, i.e., $[\eta(kT)]|_{dB}^{(ST)} \approx [\eta(kT)]|_{dB}^{(FT)}$ or, in other terms, the absolute difference $|\varepsilon(kT)|_{dB}|$ (in dB), where $\varepsilon(kT)|_{dB} \triangleq [\eta(kT)]|_{dB}^{(ST)} - [\eta(kT)]|_{dB}^{(FT)}$, is of the order of 0.1 dB or so, while, at rain onset, the fast tracker output $[\eta(kT)]|_{dB}^{(FT)}$ starts following the variation of SNR due to rain and may deviate significantly (downwards) from the slow tracker output $[\eta(kT)]|_{dB}^{(ST)}$. The beginning of a rain event (*rain detection*) is declared when the difference between the outputs of the two trackers $\varepsilon(kT)|_{dB}$ exceeds a given detection threshold, e.g. 0.3 dB. This choice sets a lower limit (sensitivity) to the detectable rain rate, which must be balanced against the false rain detection rate. An insight on what is a good choice for the detection threshold is given by Fig. 8, which shows the histogram of the probability density function of $\varepsilon(kT)|_{dB}$, the difference between ST and FT outputs (in dB), measured in dry conditions. The Gaussian fitting curve has mean 0.0049 dB and standard deviation 0.0852 dB, and a good approximation of the probability of a false alarm is given by integrating the tail of this fitting Gaussian curve for $\varepsilon > 0.3$ dB.

The "rain detector" block in Fig. 7 has the task of detecting the crossing of the above threshold. After a rain event has been detected, the rain attenuation $L_{rain}(kT)$ at sampling instants is estimated by means of

$$L_{rain}(kT) = \frac{[\eta(t_0)]|_{dB}^{(ST)}}{[\eta(kT)]|_{dB}^{(FT)}} (1-\xi) + \xi, \qquad (13)$$

where $[\eta(t_0)]|_{dB}^{(ST)}$ and $[\eta(kT)]|_{dB}^{(FT)}$ have replaced the SNR in dry and wet conditions of the original formulation (3), respectively. During a rain event, the rain flag is set *true* and the slow tracker state is "frozen" in the condition it was in when the rain started at the instant $t_0$ (see also the discussion in Sect. IV). Thus, employing the values of $L_{rain}(kT)$ provided from (13), an estimate of the rain rate is obtained according to (12). At the end of the rain event, the FT state returns in the close proximity of the "frozen" reference ST state, and when their difference falls below a further threshold (e.g., 0.1 dB), the rain detector states "end of rain" and resets the flag to *false*.

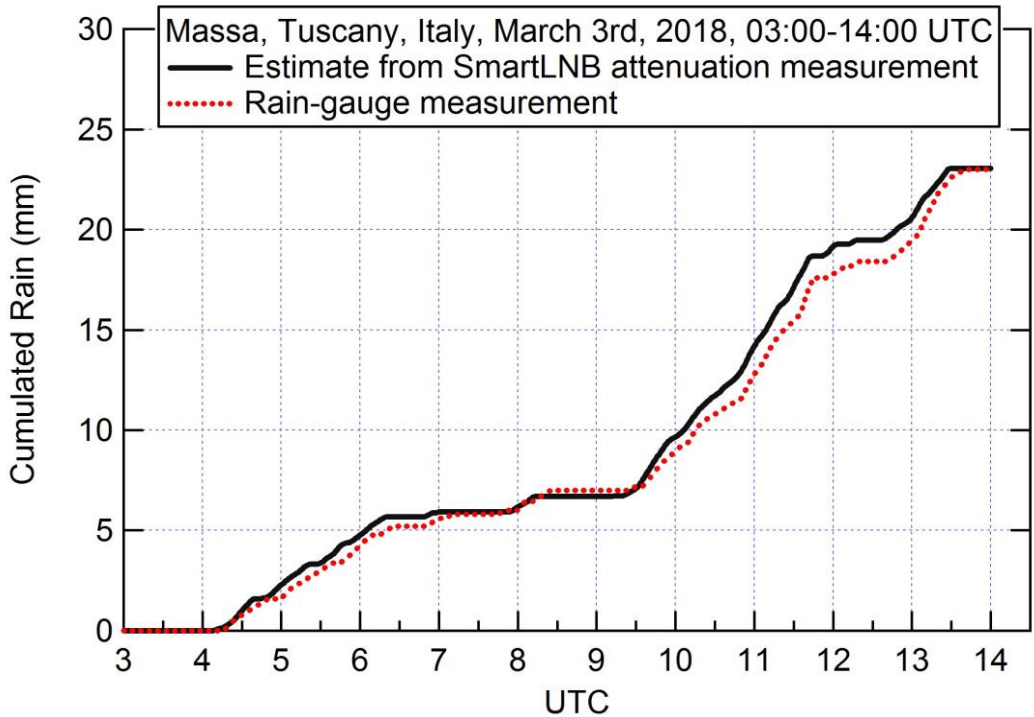

Fig. 11. Estimate of the rain rate obtained from SmartLNB measurements (solid line) made by station NEFOCAST-ITA-MS-001X (Massa, 44.0344º N, 10.1399º E), compared with conventional rain gauge measurements (dashed lines).

IX. NUMERICAL RESULTS

In order to clarify the operation of the double-Kalman-based algorithm outlined above, let us first consider the application example in Fig. 9, whose upper part shows a sequence of raw $E_s/N_0(kT)$ (light blue line) produced by a SmartLNB, along with the outputs of the two trackers (black line: $[\eta(kT)]|_{dB}^{(ST)}$; red line: $[\eta(kT)]|_{dB}^{(FT)}$). The results show that in the absence of rain (first and last segments of the plot) the outputs of the two filters are almost coincident, while during the precipitation event they depart significantly from one another. Using (13) and (12) the corresponding rain rate is then estimated, and the result is shown in the lower part of Fig. 9.

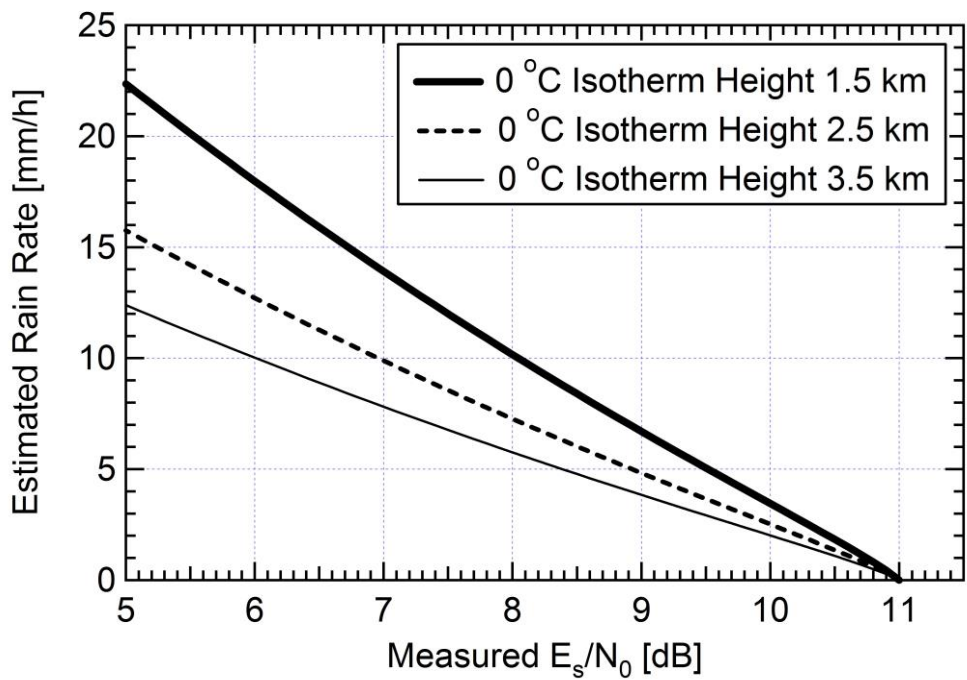

Fig. 12. Characteristics of the rain rate estimator, plotted for the typical values of the parameters of the station NEFOCAST-ITA-PI-003X (Pisa, 43.7203° N, 10.3836° E), and three different values of the 0 °C isotherm height.

To validate the accuracy of the NEFOCAST rain retrieval algorithm, Fig. 10 shows a comparison of the rainfall rate estimate (in mm/h) obtained from station ID NEFOCAST-ITA-MS-001X (Massa, 44.0344° N, 10.1399° E, elevation angle $\theta_e \cong 40°$), with the measurements provided by a nearby TBRG. Apparently, SmartLNB estimates show a fairly good agreement with the trend of data provided by the TBRG, especially as far as the position of peaks of precipitation are concerned. Furthermore, Fig. 11 plots the cumulated rain (in mm) obtained by integrating the in-stantaneous rain rates displayed in Fig. 10. Again, a good match between the curves is shown. However, it is worthwhile to point out that a perfect matching between SmartLNB and TBRG responses cannot be achieved in view of the different nature of the two sensors. As matter of fact, as illustrated in the Appendix, the SmartLNB produces a sort of quasi-instantaneous average of the signal attenuation along the wet segment of the radio path, whereas the TBRG actually measures the punctual amount of cumulated rain in a given location, so that in the latter case the rainfall rate can only be estimated by evaluating its time derivative.

Finally, of all the parameter used, the most critical one is perhaps the height $h_0$ of the 0 °C isotherm. Fig. 12 shows the characteristic curve designed to obtain the rain rate estimate from the measured SNR, once a rain event has been detected. The curve maps a value of SNR onto a corresponding estimate of rain rate according to (12) and (13), assuming the width of the melting layer is fixed at $\delta_{ML} = 500$ [20], [22]-[24]. The figure shows the results for three different values of the thermal zero height computed for the station NEFOCAST-ITA-PI-003X (Pisa, 43.7203$^o$ N, 10.3836$^o$ E). As expected, as the height $h_0$ decreases, the rain rate associated to a certain SNR increases substantially.

## X. Conclusions

We have outlined the main aspects of the NEFOCAST project, funded by the Tuscany Region, for real-time monitoring of rain fields, based on SNR measurements of opportunistic satellite signals received at a large number of SmartLNB domestic terminals scattered throughout the regional territory. Compared to the results already known in the literature, one of the key contributions of this paper is the development of a method, based on a dual slow-fast Kalman tracker, to identify the time-varying, dry-state reference value of SNR to be used both for detection of rain and for estimation of the rainfall rate. If appropriately geared, this technique is shown to effectively compensate the slow SNR fluctuations due to the apparent movement of the satellite and to remove the effect of scintillation noise. The paper also proposes a model for rain attenuation valid for a two-layer (melting and liquid) stratiform precipitation, allowing to map a measurement of the rain-induced excess attenuation onto an estimate of rainfall rate. When the thermal zero height is accurately tracked or predicted via short-term forecasts, as envisaged in NEFOCAST, the above approach leads to reliable and accurate estimates of the rainfall rate. Some experimental results have been presented to validate the proposed technique, and its performance has also been compared, with satisfactory agreement, with that of traditional tipping bucket rain gauges.

## APPENDIX

We report hereafter an analytical description and interpretation of the measures that are physically performed by a conventional TBRG and by a SmartLNB during a rain event starting at instant $t_0$, whose actual rain rate is $R_{LL}^{(Actual)}(\vec{s};t)$, where $\vec{s} \triangleq \{x, y, z\}$ is a 3D coordinate and $t > t_0$.

First, let us consider a TBRG having measurement resolution $\Delta c$ (in mm), located at a given point of coordinates $\vec{s}_0 \triangleq \{x_0, y_0, z_0\}$ and denote with $\{t_1, t_2, \ldots, t_N\}$, $t_i > t_0$, $\forall i \in \{1, 2, \ldots, N\}$, the ascending-sorted set of tip record instants. Then, within the generic $i$ th interval between two consecutive tips, i.e., for $t_{i-1} < t \leq t_i$, the measured cumulated rain is $\Delta c = \int_{t_{i-1}}^{t_i} R_{LL}^{(Actual)}(\vec{s}_0;t)\, dt$ and the estimated punctual rain rate turns out $R_{LL}^{(TBRG)}(\vec{s}_0;t) = \Delta c / (t_i - t_{i-1})$.

Now, let us denote as $\vec{\ell}$ a generic coordinate along the slanted straight path (denoted as $\Lambda$) of the radio signal across the melting and the liquid layers, i.e., from the zero-degree isothermal down to the SmartLNB. Then, the attenuation which is derived from SmartLNB measurement is related to the actual rain rate as

$$L_{\text{rain}}(t)\big|_{\text{dB}} = \int_{\Lambda} \alpha(\vec{\ell}) \left[ R_{LL}^{(Actual)}(\vec{\ell};t) \right]^{\beta(\vec{\ell})} d\vec{\ell}, \tag{14}$$

In (14), we resorted to the conventional power law, as in (6), where the coefficients $\alpha$ and $\beta$ are dependent on the coordinate $\vec{\ell}$ so as to account for the varying physical characteristics of the troposhere involved in the propagation (in our model we assumed a two-layer coarse model). The value of the rain rate $R_{LL}(t)$ obtained by solving (12) starting from attenuation (14) is thus a sort of path-averaged estimate of the actual value $R_{LL}^{(Actual)}(\vec{\ell};t)$. Also, we note that the ground projection of the slanted path $\Lambda$ has lenght $d = h_0 / \tan \theta_e$ and is aimed towards the satellite azimuth coordinate. According to the 35-year observation data from the "Modern-Era Retrospective analysis for Research and Applications" (MERRA) database [13], for a typical scenario in Tuscany, we can assume a satellite elevation $\theta_e = 40°$ and an isothermal height in the range $h_0 = 1500 \div 4000 \text{ m}$. Therefore, the path-averaged estimate provided by the SmartLNB applies to a ground segment whose lenght lies in the range $d = 1800 \div 4800 \text{ m}$.

The above discussion shows that the physical quantities actually measured by a SmartLNB and a TBRG are not coincident, so it should not be expected that the two sensors provide exactly the same outputs, even if they are in the same position.

ACKNOWLEDGMENT

The authors acknowledge the "Fondo per le Agevolazioni alla Ricerca and Fondo Aree Sottoutilizzate (FAR-FAS) 2014", agreement No. 4421.02102014.072000064

SVI.I.C.T.PRECIP. (Sviluppo di piattaforma tecnologica integrata per il controllo e la trasmissione informatica di dati sui campi precipitativi in tempo reale) of the Tuscany Region, Italy, for providing financial support of the NEFOCAST project, Greater Florence Authority for providing logistic support to the experimentation campaign and the "Pianeta Galileo" initiative of Tuscan Regional Council for its cooperation in the dissemination of the scientific results. Fruitful scientific and technical discussions with the partners of the NEFOCAST project at CNIT and CNR-IBIMET have been greatly appreciated. The authors are also indebted to A. Nerelli from Pisa and P. Lunini from Massa (both in Italy) for providing extensive rain gauge data.